\documentclass{llncs}
\usepackage{graphicx}
\usepackage{wrapfig}

\begin{document}

\title{Distributed Collections of Web Pages in the Wild}

\author{Paul Logasa Bogen II \and Frank Shipman \and Richard Furuta}
\institute{Center for the Study of Digital Libraries\\
		Department of Computer Science and Engineering\\    
		Texas A\&M University\\
		College Station, Texas 77843\\
    	\email{dcm@csdl.tamu.edu}}

\maketitle

\begin{abstract}

As the Distributed Collection Manager's work on building tools to support users
maintaining collections of changing web-based resources has progressed, questions
about the characteristics of people's collections of web pages have arisen.
Simultaneously, work in the areas of social bookmarking, social news, and
subscription-based technologies have been taking the existence, usage, and
utility of this data for granted with neither investigation into what people are
doing with their collections nor how they are trying to maintain them. In order
to address these concerns, we performed an online user study of 125 individuals
from a variety of online and offline communities, such as the reddit social news
user community and the graduate student body in our department. From this study
we were able to examine a user's needs for a system to manage their web-based
distributed collections, how their current tools affect their ability to maintain
their collections, and what the characteristics of their current practices and
problems in maintaining their web-based collections were. We also present
extensions and improvements being made to the system both in order to adapt DCM
for usage in the Ensemble project and to meet the requirements found by our user
study.

\end{abstract}

\section{Introduction}
The Distributed Collection Manager (DCM), is the successor to the Walden's Paths
Project's path maintenance utility, known as PathManager \cite{Bogen-2008}. DCM
was motivated not only by our original observations that the fluidity of web
pages leads to collections becoming stale and requiring revisions and updates
\cite{Francisco-Revilla-2001}, but also by observations that the web as a whole
was changing and that assumptions made by PathManager may no longer be valid
\cite{Bogen-2007}. Unlike PathManager, which was focused on maintaining a path in
Walden's Paths, DCM is more general and supports other forms of web-based
collections, such as bookmark lists and web resource guides.

While a path was a well-defined system artifact produced by Walden's Paths, a
general web-based collection, as DCM envisions, is a poorly-defined social
artifact. This ambiguity in what a web-based collection could be necessitates an
inquiry into what the collections that people are creating are really like.
Additionally, informal discussions with colleagues raised the question that
people may not be creating collections of web pages and are instead relying on
recollections and search to re-find previously found web pages. This raised the
question of ``What value was a system to manage collections of web pages if no
one was creating them?''. At the time PathManager was created, options, such as
social news sites, like reddit and digg, or social bookmarking like del.icio.us,
didn't exist. Now that there are options other than plain websites, bookmark
files, or recollection, are the user issues that PathManager originally attempted
to resolve still relevant? Or, do the social aspects even matter? Lastly, are
these collections purely private or do they play a social role?

Beyond these broad motivational questions there were also technical questions
that needed to be addressed. Subscription technologies, like RSS, are often seen
as a solution to a user staying updated on sites they are interested in, but do
they actually improve the problem of staying up-to-date? Does the content-only
model of RSS ignore important aspects of a page such as presentation, or
interaction? And, are subscription based collections any easier to maintain?

To understand these questions, we conducted an online survey of potential users.
From their responses we will show that people do create collections of web pages,
that they use a variety of technologies, including RSS, and that the existing
tools are inadequate. We will also show that the collections being created, even
without social technologies, often serve a social purpose. Finally, we will show
that users are primarily concerned about textual content and, possibly, imagery,
in their collections. And, that despite its focus on textual content, the lack of
intelligence in subscription aggregators makes users of subscription-based
technologies more likely to be lost in a sea of information.

Ensemble is a multi-university project funded by the NSF to add a
computing-oriented portal to the NSDL family of STEM Pathways websites. Ensemble
has a triple focus to support computing education, the application of computing
to other STEM areas, and the use of computing in science education.

One aspect of this effort is the creation of tools to support these focuses.
Since DCM is a tool to support the maintenance of collections, it provides the
Ensemble project a tool to maintain personal collections of computing
resources.

Another aspect of Ensemble is the creation of collections of web-based materials
to support the areas of focus. These collections are distributed not only in
terms of the members being distributed across the web, but the collections
themselves are spread out across the institutions collaborating on the project.
The widely distributed nature of these collections makes maintenance very
difficult. In fact, what a collection contains may be ambiguous as some
sub-collections may be maintained by communities that are not directly involved
with the Ensemble project. In response to this, DCM is being adapted to help
maintain these highly-distributed collections. 

A third aspect of Ensemble is the incorporation of a number of non-traditional
resources including a social networking sites and computing media. This
combination of traditional and non-traditional elements yields a new model of
digital library that may have unique challenges that may require a deeper
understanding of social media.

The remainder of the paper will begin with background on DCM and related works.
The fourth section of the paper will describe the method of our survey and a
summary of our respondents. Then we present our results and analysis. Finally, we
will present our conclusions and planned future work.
\section{DCM}
The Distributed Collection Manager (DCM) is being developed in order to help
maintainers of collections of found web pages monitor sites in their collections
for unexpected changes. DCM is interested in sites with an expectation to change.
Unlike sites that are expected to be static, we cannot set an threshold of change
to divide normal from abnormal behavior. Instead, a changing page has a continuum
of change where absence of change may be just as unexpected, and thus abnormal,
as a high degree of change. As a system, DCM is focused on providing a platform
that will enable future experimentation on features and analysis techniques,
while providing users with a system that augments their decision making on the
continued integrity of their collection.

In order to address these purposes, DCM was designed as a web application with a
set of supporting server-side systems linked through a common database and
repository of page versions. 

The server side consists of three parts that we have named after the three fates
of Roman mythology. Nona is the caching portion of the system. Decima is a
modular feature extraction system. Finally, Morta is an extensible analysis
system.

The front end of the system, Hannah, is a web application that allows collection
creation, collection modification, scheduling of back-end processes, and viewing
of each stage's results.

DCM supports four feature extractors. The first is a standard term frequency
count with stop word removal and stemming. Our second feature is the
Flesch-Kinkaid text readability index. The third feature is an updated version of
the Structural Algorithm \cite{Francisco-Revilla-2001}. We also support
dimensionality reduction using principal component analysis.

DCM currently provides a Kalman filter-based analysis module as described in
our previous work \cite{Bogen-2008}. Additionally, we currently have over
725,000 caches from approximately 500 websites collected over the past 4 years
in our repository.

\subsection{DCM in Ensemble}
As part of DCM's involvement with the Ensemble Project, two subsystems are
currently being developed. These subsystems are called Ananke and Ianus.

Our focus on maintaining distributed collections makes DCM well-suited to monitor
Ensemble's collections. However, unlike personal collections, Ensemble's
collection is not well-defined by a single individual. In order to support these
kind of collections we are developing a new subsystem, Ananke. This subsystem is
a crawler designed to automatically build collections for DCM to monitor complete
sites without full prior knowledge of their extent.

Additionally, unlike the semi-frequent personal attention that DCM's normal
intended usage was designed for, Ensemble needs a system with minimal attention
that only requires user intervention when a problem is detected. In order to
support this use case, we will create a second additional interface, Ianus, that
can operate in a automatic fashion on the crawled collections produced by Ananke.
When Ianus needs user intervention, it will inform the Ensemble project via
email.

Together, Ananke and Ianus will enable DCM to support large non-personal
collections in addition to smaller personal collections.Thus DCM will be able to
be deployed as a tool to help manage the Ensemble project's growing decentralized
collection of distributed resources.
\section{Related Work}
While in the past DCM has focused on the area of detecting change in web-based
collections, our questions, in this work, deal with how people currently are
creating, using, and maintaining collections of web pages. These questions deal
with some areas we have previously dealt with, such as aspects of change and
subscription technologies, and others that we have not, like bookmarking
practices and the emerging social bookmarking and news sites.

\subsection{Aspects of Change}
Previously, DCM had focused only on changes in term frequencies of the textual
content of the page \cite{Bogen-2007}. Additionally, our predecessor,
PathManager, had user structural analysis \cite{Francisco-Revilla-2001} and
context analysis \cite{Dalal-2004} is measure of the continued validity of
sites and collections.

Many techniques have been used previously to measure change of web documents.
Some projects, such as the AT\&T Internet Difference Engine, have relied on
presentation of differences using a traditional differencing algorithms
\cite{Douglis-1998}. Others, such as Zoetrope, focus on presenting a user with
changes to specific directed portions of the page \cite{Adar-2008}. Additionally,
Greenberg and Boyle used image comparison techniques to identify visual changes
between versions of web-based documents \cite{Greenberg-2006}.

Some have attempted to compile comprehensive lists of change metrics. Ivory and
Megraw identified over 150 metrics ranging from traditional text metrics to
information about styling, graphics, performance, and linkages \cite{Ivory-2005}.
Yadav \textit{et al.} identified four categories of changes: content/semantic,
presentation/cosmetic; structural; and, behavioral \cite{Yadav-2007}.

In the prior work, metrics selected were selected based on the intuition of the
researchers and not based on studies into what users do or what they actually
care about.

\subsection{Subscription Technologies}
Subscription Technologies, such as RSS and ATOM, are technologies that allow a
simplified content and metadata feed to be harvested by a system for reuse in
another context. These feeds are typically dynamicly generated so that a
retrieval of the feed always produces the latest content.

While the subscription technologies continue to be a large area of ongoing
research, including our own previous work investigating RSS as a means to
automatically augment existing paths with relevant information \cite{Dave-2004},
Liu \textit{et al.} found that while there was a large body of work about using
RSS as a resource or a tool, there was little to no work about how the readers of
RSS feeds were using them \cite{Liu-2005}. Liu delved in to topics such as how
many feeds readers read and how frequently their aggregation utilities retrieved
the feeds.

\subsection{Bookmarking Practices}
Since Vannevar Bush's \textit{As We May Think} introduced the concept of a
electronic bookmark as a coded index into a microfilm book stored inside the
Memex \cite{Bush-1945}, the concept of a bookmark has been an important component
of digital collections and hypertexts. 

Li \textit{et al.} were able to point to prior work showing that users did have
a difficulty keeping things found and organizing information. However, they did not
address how people were trying to organize information and if bookmarks were even
being used \cite{Li-2000}. Kellar \textit{et al.}'s study into how people seek
information on the Web gathered their data by collecting bookmark files and was
thus unable to give in insight into what all users were doing as opposed to what
users who used bookmarks were doing \cite{Kellar-2007}. However, the prevalence
of bookmarks has been examined three times. First, a 1998 study found that 98\%
of attendees at an academic conference focused on the Internet had bookmarks
collections \cite{Abrams-1998}. A 2001 study on how user's kept previously found
items on the web found showed that only one of their four participants used
bookmarks \cite{Jones-2001}. Lastly, a study in 2005 on members of ACM's SIGCHI
mailing lists found that 92.4\% of the participants created bookmarks
\cite{Aula-2005}. However, other work that examined actual usage of bookmarks
through click tracking \cite{Obendorf-2007} concluded that people don't revisit
bookmarks very often. This seemingly contradictory situation has not been
addressed. Why do people create bookmarks, if they are not using them?

\subsection{The Social Web}
With the rise of the Social Web came a new approach to bookmarking and news
gathering on the web. Social bookmarking and social news sites bring what were
once individual activites by a sole user, in the case of social bookmarks, or an
editor, in the case of social news, and instead allow a community to identify
interesting and relevant resources for each other. Often this involves community
voting or tagging to build these rankings. 

In the realm of social bookmarks, a large portion of the existing work has
focused on how the sites can be utilized to help inform other tasks. These range
from using social bookmarking sites to build summaries of web sites
\cite{Park-2008} to semantic web research attempting to generate ontologies from
the tags that users had applied to their bookmarks \cite{Wu-2006}.

Another major set of social bookmarking work focuses on the social aspects.
Work in this area has delved into topics like: the quality of tags
\cite{Penev-2008} and how social networks evolve \cite{Garg-2009}.

Of particular interest to our work, is prior work that attempted to answer the
questions ``Why do people create tags?'' and ``What do people use social
bookmarking cites for?'' The first question was addressed by Kathy Lee's work
examining motivations for tagging on del.icio.us \cite{Lee-2006}. In this work
the relationship between a person's tagging activity on del.icio.us and the size
of their friend list on del.icio.us. 

Much like the related social bookmarking sites, the social news sites, like
reddit, digg, and fark, consist of user found links shared amongst a community.
Unlike the social bookmarking sites, Social News sites have an emphasis on
current events and new content. Work on social news sites have been particularly
focused on the social aspect of the sites. For instance, Lerman \textit{et al.}
analyzed voting patterns on digg \cite{Lerman-2008}.

Throughout the body of work on social news and social bookmarking three questions
are not being asked. Are the collections that users are generating important to
them? Are they managing their collections? And, do social news and bookmarking
sites compliment or supplement bookmark files and subscription technologies?
\section{Methodology}
For our survey, we used a web-based survey system. We arranged our questions into
five sections. First we asked demographic information. The second section
focused on personal web-based collections. Questions were asked about who used their
collections, the tools they used, and how important their collections were to
them. The third section delved deeper in to the management of collections.
Questions were asked about the kinds of sites in their collections, the kind of
changes they care about and their experiences in maintaining these collections.
Fourth, we switched specifically to subscription technologies and their likes and
dislikes regarding them. The fifth section asks users to identify features they'd
like to see in DCM and how likely they were to use a system like DCM for
maintaining their collections.

In order to promote the survey we solicited participants through mailing lists
and social networks. In particular, we advertised on our lab's mailing list, a
departmental list for graduate students and on three social networks -- Twitter,
Facebook, and reddit. The survey was conducted over a two week period in December
2009.

\subsection{Demographics}
We received 125 responses for the survey. 41.6\% of the respondents were
undergraduate students, 28\% were graduate students, while the remaining 30.4\%
were not students. Ages of respondents ranged from 18 to 52 with the average age
of respondents being 25.27. 80 users came from a computing and information
sciences background. 12 from a science background, 10 from a liberal arts and
social science background, 8 from engineering, and 1 from education. respondents
came from a wide range of localities. North America comprised the majority with 
with 75 respondents. Additionally, we had 19 Europeans, 6 from Australia and
New Zealand, 6 Asians, 2 Middle Easterners and 2 South Americans respond.
\section{Results}
As discussed previously, we asked questions in roughly areas: collection usage,
management techniques, subscription technologies, and desired features.
Statistical analysis were performed using R and gretl. All probabilities, unless
otherwise noted, were results of n-way analysis of variance using a linear model
with factor interaction accounted for.

\subsection{Collection Usage}
Several questions asked by our survey focused on the usage of collections of web
pages. The first question was if they had collections of web-pages. 45.6\% of
respondents reporting having a collection of web sites. However, an additional
15.2\% indicated later in the survey that they did maintain a collection when
specific when we asked about more specfic kinds of collections, totaling 60.8\%.
Of those who have collections, only 4.5\% reported that they never
revisit their collections, while 80.3\% revisit their collections daily.

The next question was if collections were private or if they were shared. Only
22.81\% of the respondents indicated that someone other than themselves used
their collections. 53.85\% of respondents who shared their collections of web
sites did so with family. These respondents created collections that tended
to change more often than collections created by people not sharing with their
family members ($p=0.05$). 23.08\% of those who shared, were sharing their
collections with friends. They tended to lose track of their collections more
often than respondents who weren't sharing with friends ($p=0.08$). 69.23\%
indicated small groups of people either in organizations, a work environment, or
in a academic project group. There respondents created more frequently changing
collections than people whose collections were not being used by a group
($p=0.07$). Likewise people who created both collections that were used by their
family and in a professional/academic setting tended to have the most frequently
changing collections ($p=0.04$).

When we asked what type of sites people were interested in for their collections,
social news sites and traditional news sites dominate the kind of sites that
respondents keep in their collection of web sites. Comics come in at third while
blogs and social networks were cited the fourth and fifth most frequently.

\begin{wrapfigure}{l}{.55\textwidth}
	\begin{center}
		\includegraphics[width=.54\textwidth]{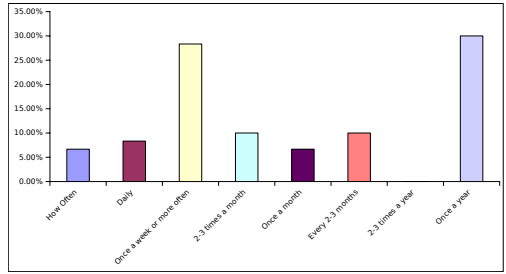}
	\end{center}
	\caption{Relative frequency in which users lose track of their collections.}
	\label{losing_track}
\end{wrapfigure}

\subsection{Collection Management Techniques}
Another area of interest was what tools people were using to maintain their
collection. Every respondent except for one reported using some sort of tool for
maintaining their personal collections. Traditional bookmark usage was common,
with 85.45\% of respondents using them. However, despite the fact that the
majority of respondents (57.14\%) were consumers of social news and bookmarking
sites, only 23.64\% of respondents were actually using social news or social
bookmarking sites to maintain their collections. 12.73\% of respondents were
using a subscription technology like RSS and 10.91\% were using other kinds of
web pages (like Wikis or hand-written HTML) to maintain their collection. We
found that for certain factors, the kind of tool was a statistically significant
detriment to the respondent using the tool. Respondents using bookmarks found it
more difficult to maintain their collections than respondents who didn't
($p=0.02$). Respondents using no tools ($p=0.03$), their history mechanism
($p=0.08$), or their email ($p=0.08$) to maintain their collections perceived
them changing more dramatically than others.	

Of the users who used a subscription-based technology, all of them also used
bookmarks, and 14.29\% of them also used some sort of web site. 52.17\% of
bookmark users used another technology.

For types of change our results appear contradictory to speculations made b by
others in the literature. Content changes made up the vast majority of changes
people were interested in. 89.5\% of respondents indicated ``content'' as an
aspect of change they were interested in. The second-highest aspect was ``visual"
with only 5.08\% interested. However, we do suspect that some of ``content'' as
defined by the respondents still included imagery, particularly since comic sites
showed such a frequent occurrence in respondent collections.

When respondents were asked ``How often would you say that you lose track of
sites in your collection?'' Respondents were given the options of daily, once a
week or more often, ``2-3 times a month'', ``once a month'', ``every 2-3
months'', ``2-3 times a year'', ``once a year'', ``rarely'' or ``never''. As
figure \ref{losing_track} shows, we found a bimodal distribution with means at
``2-3 times a month'' and ``never''. However, we were not able to correlate the
bimodality of our results to any data we collected.

We performed a Pearson's coefficient calculation between each pair of questions.
From these coefficients we were able to find a number of correlations between
factors dealing with collections. People who create work collections were found
to have less dramatic changes than other kinds of collections ($p=0.06$). The
more important a collection was to a respondent, the more time they spend
maintaining it ($p=0.09$) and the more difficulty they had in keeping track of it
($p=0.11$). Collections that were revisited more often were also more difficult
to maintain ($p=0.10$). Difficult to maintain collections took more time to
maintain ($p<0.01$). Subscription-based collections took more time to maintain
than non-subscription technologies ($p=0.02$).

\subsection{Subscription Technologies}
When respondents were asked what they liked and disliked about subscription
technologies, 86.2\% of respondents had the same like -- consolidation of several
sites content in to one easy, quick place to read everything. However, four major
kinds of dislikes were found. 37.5\% of them said that the pace of updates caused
information overload and that they need some kind of filtering method. 33.3\%
complained that the subscription feeds were often only a subset of the content of
the site. Some feeds would miss items, some wouldn't have consistent metadata,
others wouldn't have the entire article text, and some wouldn't provide locations
of relevant images. 12.5\% found the selection of sites to be publishing feeds to
be sub-par or limited and finally, 8.3\% found the interfaces of the readers
themselves to be inadequate.

\subsection{Desired Features}
Finally, we asked users what features they were interested in for a system for
managing their collections of web pages. 36 users provided substantive answers.
Of those 36, 14 indicated various social web features like sharing, voting,
tagging, and recommendation. 12 indicated that they wanted a system that was easy and
simple. 7 users wanted to be automatically informed of updates, 6 wanted
categorization, 5 wanted to be able to define filters or priorities to limit
information from sources they were less interested in, 4 wanted to be able to
easily view collection members from inside the system. 
\section{Revisiting DCM}
\begin{figure}
	\begin{center}
		\includegraphics[width=.75\textwidth]{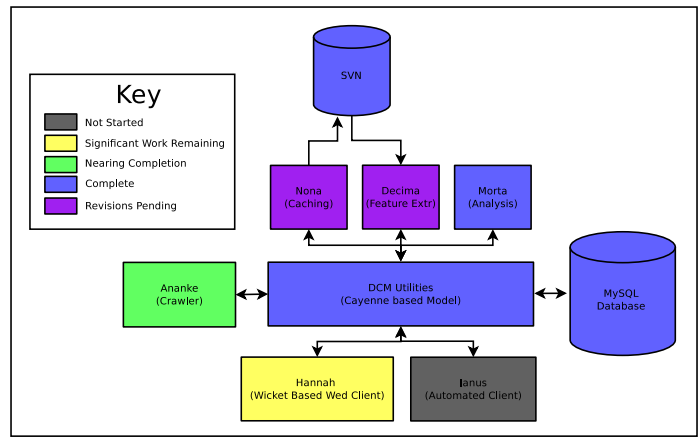}
	\end{center}
	\caption{Current status of Distributed Collection Manager.}
	\label{system}
\end{figure}

From these results we have begun to extend DCM. This means we need to both create
new modules and subsystems and extend current capabilities. Our current status
can be seen in figure \ref{system}. Nona was originally intended to only cache
html pages. However, many of respondents were interested in monitoring webcomics
and to mix feeds with traditional websites. Therefore, Nona needs to be extended
to start gathering images and be able to pull utilize feeds. Naturally, new types
of data being retrieved means that we need to be able extract features from them.
Image-centric methods may comprise image differencing algorithms, fingerprint
generation, or measures of visual characteristics such as color usage,
brightness, or saturation. For the feeds, the availability of author-supplied
meta data provides a source of features that would be difficult or impossible to
extract from traditional HTML materials. Currently collections are organized in a
list structure. However, our respondents' desire for categorization and tagging
requires replacing the simple list with a tree structure. Additional usability
improvements in the areas of browsing and editing of collections are also needed
to help meet respondent interest in ease of use. Finally, the implementation of a
feedback system, so that users can train Morta for what they view as normal
changes, is needed to provide customizability desired by respondents.
\section{Conclusions and Future Work}
With the continued existence of web-based collections confirmed and their usage
identified as primarily personal, the basic motivations of DCM are validated.
Likewise the kinds of sites we are currently analyzing, blogs and news sites, are
sites of high interest. Additionally, social networks, social news sites, and
comics are sites of interest that deserve our attention. The inclusion of comics
indicates that some measure of change of pertinent imagery would be of interest
to potential users.

Additionally, since DCM is aimed at finding unexpected changes, it may be useful
in helping cut through the information overload experienced by users of
subscription-based technologies. In order to support these users, we intend on
extending DCM with the capability to montior these feeds.

Third, the indication that poor interfaces are a common problem with current
aggregators suggests that further study in to the shortcomings of aggregator
interfaces may be warranted to try and avoid the mistakes others have made.

Beyond the results of our study, in progress work on building a ground-truth
collection of page changes to evaluate our methods will be continued. Once this
ground-truth is established, we will evaluate the effectiveness of different
features in web pages and the suitability of our Kalman Filter based analysis
mechanism. 

Finally, a follow-up user study is currently being conducted. This study requests
users to submit personal collections, such as bookmark files, for us to utilize
not only as collection in DCM, but also to help us to gain further insight in to
the characterization of user collections.
\section{Acknowledgements}
The Distributed Collection Manager is being developed as part of the Ensemble
Computing Pathways Project. Ensemble is funded by the National Science
Foundation, NSDL program, DUE 0840713.

\bibliographystyle{splncs03}
\bibliography{references}

\end{document}